\documentclass[%
 reprint,
 amsmath,amssymb,
aps,
pra,
superscriptaddress
]{revtex4-2}

\usepackage[utf8]{inputenc}
\usepackage[T1]{fontenc}
\usepackage[greek,english]{babel}
\usepackage{textcomp}
\usepackage{physics}
\usepackage{amsmath}
\usepackage{graphicx}
\usepackage{diagbox}

\usepackage{subfigure}
\usepackage[justification=justified]{caption}
\usepackage{dcolumn}
\usepackage{bm}
\usepackage{siunitx}
\usepackage{hyperref}
\usepackage{pgfplots}
\pgfplotsset{compat=newest}
\usepgfplotslibrary{groupplots}
\usepgfplotslibrary{dateplot}
\usetikzlibrary{spy}

\begin{document}

\preprint{PRA/123-QED}

\title{Detecting a logarithmic nonlinearity in the Schrödinger equation\\ using Bose-Einstein condensates}

\author{Sascha Vowe}
 \email{vowe@physik.hu-berlin.de}
 \affiliation{%
 Humboldt-Universität zu Berlin\\
 Newtonstraße 15, 12489 Berlin, Germany
}%

\author{Claus Lämmerzahl}%
 \affiliation{%
 Center of Applied Space Technology and Microgravity (ZARM), Universität Bremen\\
 Am Fallturm, 28359 Bremen, Germany.
 }%
\author{Markus Krutzik}%
\affiliation{%
 Humboldt-Universität zu Berlin\\
 Newtonstraße 15, 12489 Berlin, Germany
}%

\date{\today}

\begin{abstract}
We study the effect of a logarithmic nonlinearity in the Schrödinger equation (SE) on the dynamics of a freely expanding Bose-Einstein condensate (BEC). 
The logarithmic nonlinearity was one of the first proposed nonlinear extensions to the SE which emphasized the conservation of important physical properties of the linear theory, e.g.: the separability of noninteracting states.
Using this separability, we incorporate it into the description of a BEC obeying a logarithmic Gross-Pittaevskii equation.
We investigate the dynamics of such BECs using variational and numerical methods and find that, using experimental techniques like delta kick collimation, experiments with extended free-fall times as available on microgravity platforms could be able to lower the bound on the strength of the logarithmic nonlinearity by at least one order of magnitude.

\end{abstract}

\maketitle


\section{\label{sec:introduction}Introduction}
Quantum theory is the most fundamental theory in physics and on the elementary level, all types of matter, as well as radiation, have to be described by it.
To this day, the time evolution of quantum systems as predicted by the Schrödinger equation (SE) has been confirmed in many experiments \cite{Lamoreaux1992,Arndt2014}. 
Nonetheless, whether the SE can be regarded as a complete description or rather a linearized approximation of a more general theory is still an open question since, despite its great success, quantum theory has a few unresolved problems, e.g. its missing connection to General Relativity and the measurement problem.
\\\indent
However, it is not obvious how to modify the SE in order to tackle these problems. Employed modifications are for example the generalization of the uncertainty relation \cite{Maggiore1994,Pikovskietal2012,Rudnicki2016}, the addition of nonlinear \cite{Weinberg1989} and stochastic \cite{GRW1986} terms or higher derivatives \cite{LaemmerzahlBorde01} to the Schrödinger equation.
Such modifications can be the result of a theory of quantum gravity in its low-energy limit (see e.g. \cite{AlfaroMoralesTecotlUrrutia02a}), or may result from attempts to find a solution to the measurement problem \cite{Penrose1996}.
Current research focuses more on stochastic nonlinear terms that describe the wave function collapse or decoherence, which can be either induced spontaneously or by gravity \cite{Bassi2013, Bassi2017}. 
A deterministic nonlinear time evolution of the wave function is nowadays rarely considered,
then usually by the inclusion of a semi-classical description of gravity as in the Schrödinger-Newton equation \cite{Diosi1984}.
This can be mainly contributed to the fact that the most prominent generalization of a nonlinear SE, as described by Weinberg \cite{Weinberg1989}, leads to problems of locality if extended to the case of multiple entangled particles \cite{Gisin1990,Czachor1991,Polchinski1991}.
Even though it is conjectured that this can be concluded for all nonlinear deterministic extensions \cite{Gisin1995,MIELNIK2001}, there have also been subsequent attempts to extend existing deterministic nonlinear SEs to the case of multiple particles that do not violate locality \cite{Czachor1997, Czachor2002} or have disputed the apparent violation of relativity by nonlinear quantum mechanics at all \cite{Polchinski1991,Kent2005}.

Two special cases of nonlinearity have received wider attention.
One is the first formulation of a fundamentally nonlinear SE which emphasized the necessary separability of non-interacting states: the logarithmic Schrödinger equation (LogSE) proposed by Bialynicki-Birula and Mycielski \cite{Bialynicki-Birula1976}.
The nonlinear time evolution is described by
\begin{align}
    i\hbar\frac{\partial}{\partial t} \Psi = \Big(-\frac{\hbar^2}{2m}\laplacian + V 
    -b\ln \alpha\abs{\Psi}^2 \Big) \Psi . \label{eq:logarithmic-schroedinger-equation}
\end{align}

The LogSE comprises terms of the ordinary SE and the nonlinear term $-b\ln \alpha\abs{\Psi}^2$, where $\alpha$ is a physically irrelevant real constant (as it only leads to a global energy shift) of dimension $L^3$ and $b$ the strength of the logarithmic nonlinearity in units of energy.
It keeps important properties of the linear SE, e.g.: conservation of probability and norm, invariance under permutations, Galileo transforms and more. Most importantly, it guarantees that the time evolution of product states can be separated for all times.

The second case is the Gross-Pitaevskii equation
\begin{align}
    i\hbar\frac{\partial}{\partial t} \Psi  = \Big(-\frac{\hbar^2}{2m}\laplacian + V
 +g\abs{\Psi}^2 \Big) \Psi,
 \label{eq:gp-schroedinger-equation}
\end{align}
which describes the dynamics of a Bose-Einstein condensate (BEC). Its nonlinearity is just an effective description of scattering processes taking place in the degenerate quantum gas.
As macroscopically sized quantum objects, BECs have proven to be suitable candidates for testing fundamental physics, whether it be the Schrödinger equation, General Relativity and their possible intersections \cite{Herrmann2010,Raetzel_2018,Howl2019}.
These dedicated tests are important, as they broaden the domain of application of quantum theory or might give hints what direction to follow in search for deviations.
This is especially true with regard to recent developments in long-fall-time experiments of BECs in microgravity \cite{vanZoest2010, Rudolph2011, Muntinga2013} on sounding rockets \cite{Becker2018} and in space \cite{Aguilera2014, Elliott2018, beccal2019}. 

In this article we study deviations due to a possible fundamental logarithmic nonlinearity in the time evolution of Bose-Einstein Condensates (BECs). 

The current upper bound on the strength of the logarithmic nonlinearity stems from neutron optical diffraction experiments \cite{Gahler1981} in which a neutron wave packet's lateral evolution of $\abs{\Psi (\Vec{r},t)}^2$ diffracted on a straight edge was observed. They found that $b\leq\SI{3.3 e-15}{eV}$.

The article is structured as follows:
In section 2 we derive an equation describing a BEC at zero temperature including the logarithmic nonlinearity. 
In Section 3 we investigate the dynamics of the proposed equations describing the BEC. 
For this, we use numerical simulations of feasible experiments that could be able to establish new upper bounds on the nonlinearity's strength. 
In section 4 we analyze possible shortcomings of our investigation and proposed experiments with their respective error budgets.

\section{Logarithmic Gross-Pitaevskii equation}

The logarithmic nonlinearitiy can be incorporated into the theoretical description of a BEC by starting with basic assumptions as made in many textbook examples on Bose-Einstein condensation.

We assume that the condensates contains a large number $N\gg 1$ of bosons, so that we can approximate the field operator as a wave function $\hat{\psi}(\{\Vec{r}\},t) \approx \psi(\{\Vec{r}\},t)$, where the set of position vectors $\Vec{r}_1, \Vec{r_2}$...$\Vec{r}_N$ is written as $\{\vec{r}\}$.
Furthermore, we assume that all $N$ bosons occupy the same ground state $\psi$ and are not correlated with each other. Thus, we can write the wave function in a Hartree ansatz as $\Psi(\{\Vec{r}\},t)\approx \bigotimes_{i=1}^N \psi (\Vec{r}_i,t)$ with $\int \abs{\Psi(\{\Vec{r}\},t)}^2\mathrm{d} \{\Vec{r}\}= N$. 
In the low energy limit, the scatter process of particle $i$ with particle $j$ in the dilute gas ($\rho \ll a^{-3}$, where $a$ is the s-scatter length) is described by the approximated binary interaction potential $V(\Vec{r}_i,\Vec{r}_j)=g\delta (\Vec{r}_i-\Vec{r}_j)$, with $g=\frac{4\pi \hbar^2 a}{m}$.
\\
\indent
The Hartree ansatz enables us to take advantage of the separability property of the logarithmic nonlinearity. Using this, we obtain
\begin{equation}
    b\ln \abs{\Psi (\{\Vec{r}\},t)}^2 = b \sum_{i=1}^N \ln \abs{\psi (\Vec{r}_i,t)}^2.
\end{equation}
This makes the logarithmic nonlinearity part of the single particle Hamiltonian. The Hamiltonian is then
\begin{align}
     H =&
    \sum_{i=1}^N 
    \bigg[-\frac{\hbar^2 \nabla^2_{\vec{r}_i}}{2m}
    +V(\vec{r}_i,t) 
    -b \ln{|{\psi(\vec{r}_i,t)}|^2}
    \bigg] \nonumber
    \\
    +& 
    g \sum_{i < j}^N 
    \delta(\vec{r}_i-\vec{r_j}).
\end{align}
Imposing the stationary condition 
\begin{align}
    \delta \int L \mathrm{d}t &=  0 \label{eq:stationary-condition}\\
    &=\delta 
    \bigg[
    \int
    \frac{i\hbar}{2} \Big(
    \Psi^\dagger \frac{\partial}{\partial t} \Psi - \Psi \frac{\partial}{\partial t} \Psi^\dagger
    \Big)
    \mathrm{d}\vec{r}\mathrm{d}t + \int E\mathrm{d}t
    \bigg] , \nonumber
\end{align}
where $E= \expval{H}{\Psi}$, one can derive the equation governing the time evolution which is
\begin{align}
     i\hbar \frac{\partial\Psi(\vec{r},t)}{\partial t} 
     =& \frac{\delta E}{\delta \Psi^\dagger}
     \\
     =&
     \bigg[
    -\frac{\hbar^2}{2m} \nabla^2
    + V(\vec{r},t) 
        \bigg] 
    \Psi(\vec{r},t)
    \nonumber\\
    +&\bigg[
    - b \ln{|\Psi(\vec{r},t)|^2}
    + g |\Psi(\vec{r},t)|^2
    \bigg] 
    \Psi(\vec{r},t).
    \label{eq:LogPGE}
\end{align}

We absorbed $N$ into the wave function $\psi \xrightarrow{} \sqrt{N}\Psi$ and approximate $N^2\approx N(N-1)$ during calculation of the energy functional $E$.
The resulting equation \eqref{eq:LogPGE} is constituent of the usual linear kinetic and potential energy operators, the nonlinear terms from the Gross-Pitaevskii interaction, and logarithmic nonlinearity. We hence refer to it as the logarithmic Gross-Pitaevskii equation (LogGPE).

The LogGPE preserves all properties associated with density-dependent nonlinearities such as conservation of probability and invariance under permutation. The separability is lost due to the interaction term. Note that any nonlinearity, which is homogeneous (as required by Weinberg) or otherwise obeys the separability condition, can be incorporated in the same way as done here.
\begin{figure*}[htb]
    \centering
    \begin{minipage}[t]{0.49\textwidth}
        \includegraphics{{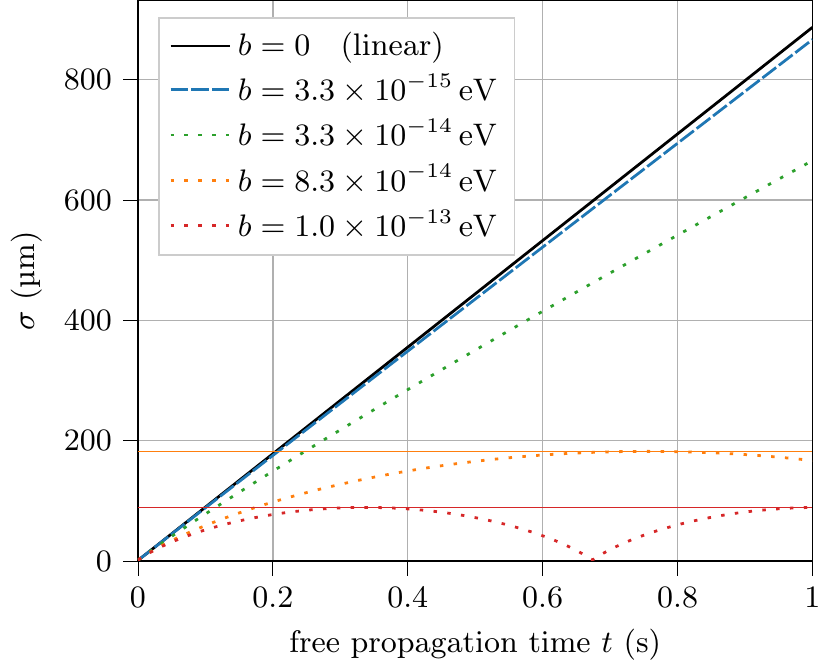}}
    \end{minipage}
    \begin{minipage}[t]{0.49\textwidth}
        \includegraphics{{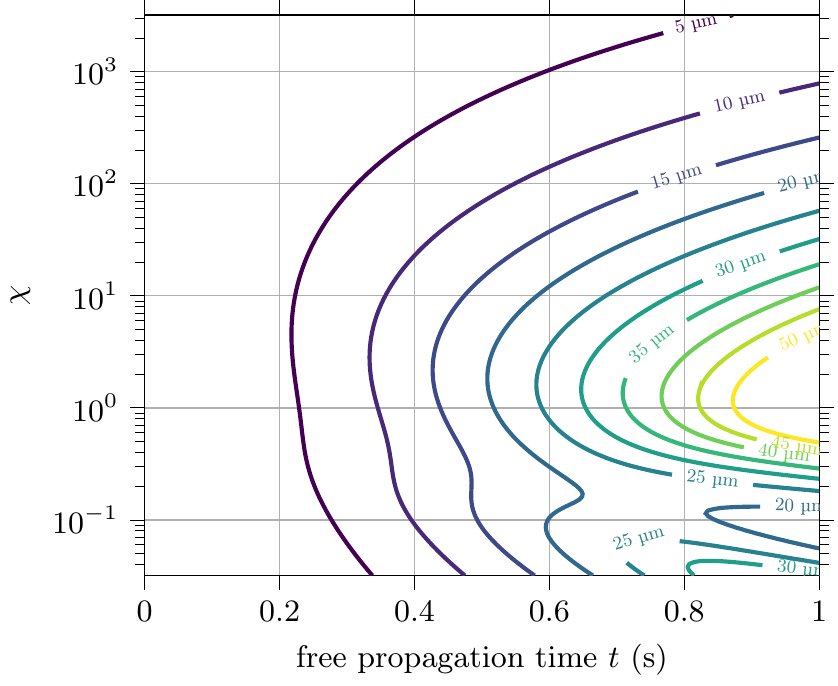}}
    \end{minipage}
    \caption{\textbf{Left}: Numerically simulated time evolution of a spherically symmetric BEC's width $\sigma(t)$ during free propagation for different values of $b$. The blue (dashed) curve represents the evolution under the current known upper bound on $b$. The two horizontal lines show the calculated maximum width from equation \eqref{eq:maximum-width} for the two examples for which spatial confinement is visible. \textbf{Right}: Contour lines depicting the difference in width of linear and nonlinear expansion $\sigma(t;b=0)-\sigma(t;b=\SI{3.3e-15}{eV})$ for different $\chi$ (see eq. \eqref{eq:chi-def}). The parameter $\chi$ has been varied by changing the initial width $\sigma(0)$ while keeping $N=\SI{5e4}{}$, $a=90a_0$ and $\Dot{\sigma}(0)=0$. For example, $\chi=10^3$ corresponds to roughly \SI{1}{\micro\meter} and $\chi=10^{-1}$ to \SI{30}{\micro\meter} initial width of the simulated BEC.}
\end{figure*}
\subsection{Variational Solutions of the LogGPE}
Finding analytical solutions to nonlinear problems is rather difficult, therefore we retain to discuss the dynamics using variational methods and the system's Lagrangian.

We can use equation \eqref{eq:stationary-condition} to obtain an approximated result for the time evolution of a logarithmic BEC by making an ansatz for the wave function as described in \cite{Perez1996,Perez1997}. 
We will assume a Gaussian shaped wave function
\begin{align}
    \Psi &(x,y,z,t) =  \prod_{\eta=x,y,z} \frac{1}{\sqrt{2\pi\sigma_\eta^{2}(t)}} \nonumber\\
    \times
    &\exp (
    -\frac{(\eta-\eta_0 (t))^2}{4\sigma_\eta^2(t)} 
    +i \eta \alpha_\eta (t)
    +i \eta^2 \beta_\eta (t)
    )
    ,
    \label{eq;gaussian-ansatz}
\end{align}
with phase terms $\alpha_\eta$ and $\beta_\eta$, which are proportional to the average velocity and inverse radius of curvature respectively.
The gaussian enevlope is chosen as it keeps its shape in the linear limit and is the soliton solution of the LogSE.
Using this ansatz for the wave function and setting 
$V(\vec{r},t) = m/2 (\omega_x^2 x^2 + \omega_y^2 y^2 + \omega_z^2 z^2) $ 
one can solve the corresponding Euler-Lagrange equations for the Lagrangian $L(t, \eta_0 (t), \alpha_\eta (t), \beta_\eta (t))$. 
From the resulting equations we obtain a set of coupled ordinary differential equations describing the time evolution of the Gaussian trial function's width under the influence of dispersion, the harmonic potential, and both nonlinearities:
\begin{align}
       & \frac{\partial^2}{\partial t^2}  \sigma_x (t) =
    \frac{\hbar^2}{4m^2}\sigma_x^{-3}(t)
    - \omega_x^2 \sigma_x (t)\nonumber\\ 
    & + \frac{\hbar^2 N a}{4 m^2 \sqrt{\pi}} \sigma_x^{-2}(t) \sigma_y^{-1}(t) \sigma_z^{-1}(t)
    - \frac{b}{m} \sigma_x^{-1}(t). \label{eq:gaussian-width-over-time}
\end{align}
The equations for $\sigma_y (t)$ and $\sigma_z (t)$ can be obtained by cyclic permutation $x\xrightarrow{} y\xrightarrow{} z \xrightarrow{} x$. From equation \eqref{eq:gaussian-width-over-time} we see that the coupling between spatial dimensions stems from the Gross-Pittaevskii interaction. The logarithmic nonlinearity  does not induce a coupling between different spatial dimensions due to its separability and the Hartree ansatz we made.
The applied approximations, namely binary interaction potential and separable wave function, require us to search for deviations in regimes where the gas is dilute and atomic interactions weak.
In addition, the nonlinearity's contribution to the overall energy of a trapped BEC is very small in presence of two-body interactions.
Experiments searching for deviations with trapped BECs are hence not suitable, even though many features, e.g. instability conditions, are altered by the logarithmic nonlinearity.
Therefore, we will investigate deviations in the ballistic expansion of BECs.

\section{Dynamics of a logarithmic Bose-Einstein Condensate}
In this section we study the dynamics of a freely expanding logarithmic BEC released off a trap under typical experimental parameters. 
We look at the time evolution of the BEC's width after release of a spherically symmetric trap (so that $\sigma_x(t)=\sigma_y(t)=\sigma_z(t) =\sigma (t) $) and show how the width as a function of time differs due to the logarithmic nonlinearity's influence.
This idea is very much alongside the lines of the last experimental tests using neutron-optical diffraction experiments in 1981 \cite{Gahler1981}.
Then, we discuss the application of delta kick collimation (DKC) to magnify these effects and follow up with an error estimation of the discussed experiment.

\subsection{Free expansion of a spherically symmetric BEC} \label{sec:a}

When the LogSE was first proposed it turned out especially appealing because it allowed for Gaussian shaped, solitonic solutions with width $\sigma_\mathrm{eq} $. 
If the wave function is of width $\sigma_\mathrm{eq} $, linear dispersion and nonlinear self-interaction are in equilibrium and $\abs{\Psi(\vec{r},t)}^2$ does not change its appearance. 
In contrast to the linear time evolution, it leads to a spatial confinement of the wave function.
Thus, we are searching for an unexpected narrowing of the matter wave packet.
For the spherically symmetric case, the three equations of~\eqref{eq:gaussian-width-over-time} reduce to
\begin{align}
            \frac{\partial^2}{\partial t^2}  \sigma (t) &=
    \frac{\hbar^2}{4m^2}\sigma^{-3}(t)
    - \omega^2 \sigma (t)\nonumber\\ 
    & + \frac{\hbar^2 N a}{4 m^2 \sqrt{\pi}} \sigma^{-4}(t) 
    - \frac{b}{m} \sigma^{-1}(t).
    \label{eq:spherically-symmetric-width-over-time}
\end{align}

To explore the logarithmic nonlinearity's influence on a BEC, we numerically integrate equation \eqref{eq:spherically-symmetric-width-over-time} in order to obtain the width's time evolution $\sigma (t)$. 
We employ typical experimental parameters of a $^{87}$Rb condensate ($N=\SI{5e4}{}$, $a=90a_0$, $\sigma (0)=\SI{2.5}{\micro\meter}$, $\Dot{\sigma}(0)=0$) and a free propagation time of \SI{1}{s}. 
The left graph of fig. 1 depicts $\sigma (t)$ of this BEC for different values of $b$.

One can see a narrowing of the BEC's width over time compared to the usual GP energy driven expansion due to the logarithmic nonlinearity. 
The blue (dashed) line corresponds to the current known upper bound $b<\SI{3.3e-15}{eV}$. 
After $\SI{1}{s}$ of free propagation it differs by $\approx\SI{20}{\micro\meter}$ compared to the linear case ($b=0$). 
The dotted (green, orange, red) lines depict the expansion for larger values of $b$.
We chose $b$ arbitrarily, so one can see the spatial confinement of the wave function taking place.
One can see that there exists a maximum width $\sigma_\mathrm{max}=\max \sigma (t)$ which is, as we will show, not only a function of $b$, but also the initial conditions of the BEC ($\sigma(0)$, $\Dot{\sigma}(0)$, $N$ and $a$). 

The logarithmic BEC's maximum width can calculated using the energy functional $ \expval{H}{\Psi}$. Inserting the Gaussian ansatz from equation \eqref{eq;gaussian-ansatz},
the energy per particle is
\begin{align}
        \frac{E(\sigma (t))}{N} =
      \frac {3}{2} m \Dot{\sigma}^2(t)
    +    {\frac {3{\hbar }^{2}}{8m{\sigma}^{2}(t)}}+\frac{3}{2}\,m{\omega}^{2}{\sigma}^{2}(t) \nonumber\\
    +{\frac {\hbar^2 N a}{4\sqrt{\pi}m{\sigma}^{3}(t)}}
    +3b \ln  \sigma(t) + C,
    \label{eq:energy-per-particle}
\end{align}
where the time derivative joined the equation via the relation $ m\Dot{\sigma}(t)=-2\hbar \beta (t) \sigma (t)$, which is obtained by solving the Euler-Lagrange equations, the center of mass motion was set $\alpha_\eta(t)=0$, and $C$ is an arbitrary constant.
As the BEC is released ($\omega=0$), the width immediately starts to increase. 
Hence, energy contributions with $\sigma(t)$ in the denominator become smaller.
Due to its conservation, the initial energy is transfers to the $\Dot{\sigma}(t)^2$ dependent term and, in the case of $b=0$, reaches a constant value in the far field. 
If $b\neq 0$, the energy of the logarithmic constituent is increasing and will ultimately counter the dispersion. 
This leads to a contraction until, again, the pressure of Heisenberg uncertainty and two-particle interactions lead to an expansion.\\
\indent
Asserting that the initial energy $E(\sigma(0))$ completely transfers into the logarithmic energy contribution (at which point the BEC can only contract), we find the formula for the maximum value of $\sigma(t)$ is
\begin{align}
    \sigma_\mathrm{max} = \max\sigma (t) = \sigma(0)  \exp \chi,\label{eq:maximum-width}
\end{align}
where
\begin{align}
    \chi =
    \frac{1}{3b}\bigg(
     \frac {3}{2} m \Dot{\sigma}^2(0)
    +\frac{3\hbar^2}{8m\sigma^2(0)}
    +\frac{\hbar^2Na}{4\sqrt{\pi}m\sigma^3(0)}
    \bigg).
    \label{eq:chi-def}
\end{align}

This maximum width is indicated in the left graph of figure 1 by horizontal lines for the two examples for which the confinement is visible. 
The dimensionless parameter $\chi$ is the ratio of initial energy to $b$. 
\\
\indent
The right graph of fig. 1 depicts the contour lines of the difference in width of linear and nonlinear free expansion $\sigma(t;b=0)-\sigma(t;b=\SI{3.3e-15}{eV})$ for different $\chi$. 
Since $b$ is set, we change the initial width $\sigma(0)$ for variation of $\chi$. 
The values for $N$,  $a$ and $\Dot{\sigma}(t)$ are the same as in the simulations of fig. 1-Left.
We see that in the regime of large $\chi$,  smaller absolute deviations of $\sigma(t)$ are found. 
This is due to the prevalence of interaction and kinetic energy in the system. 
In the vicinity of $\chi\approx 1$, we find the strongest deviations. 
This is the region in which we find equilibrium of logarithmic self-interaction and dispersion which results in solitonic behavior of the logarithmic BEC. 
When $\chi \ll 1$, the initial energy of the system consists mainly of logarithmic self-interaction energy
and the initial width is the upper bound $\sigma_\mathrm{max}\approx \sigma(0)$.
In these cases the BEC does not even disperse, but immediately starts to contract  until the contraction is countered by the resulting repulsive two-body interactions.
\begin{figure*}[htb]
    \centering
    \begin{minipage}[t]{0.49\textwidth}
        \includegraphics{{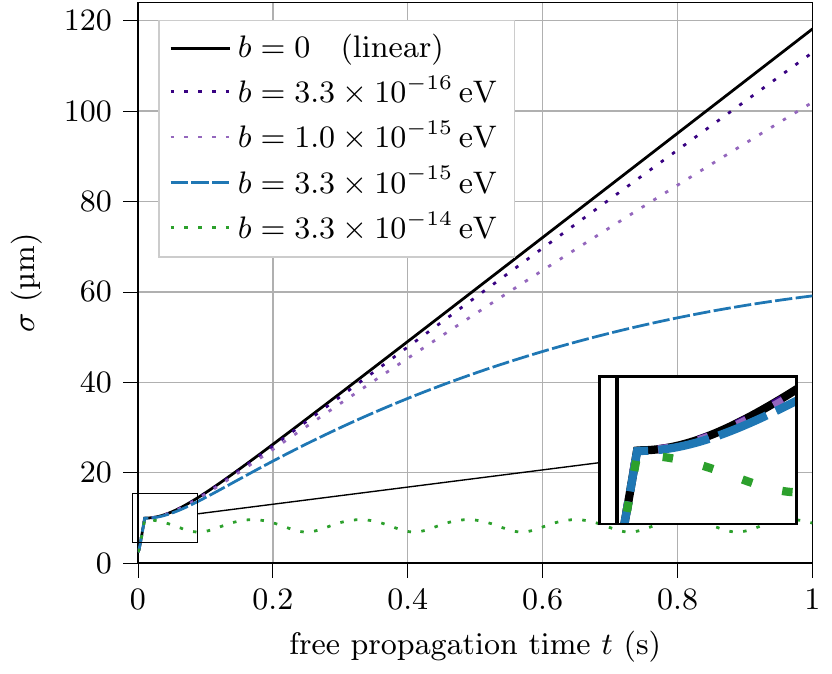}}
    \end{minipage}
    \begin{minipage}[t]{0.49\textwidth}
        \includegraphics{{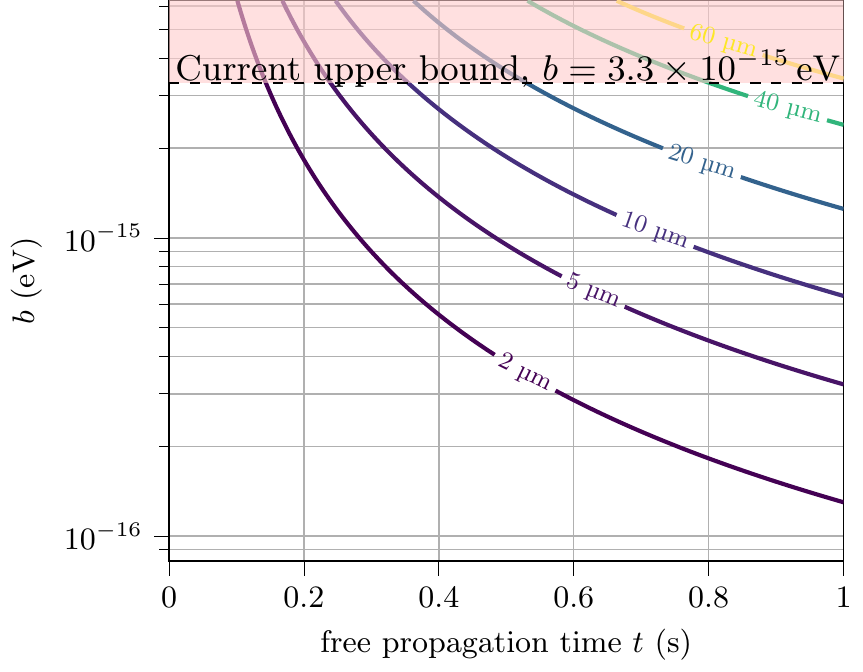}}
    \end{minipage}
    \caption{\textbf{Left}: Numerically simulated time evolution of $\sigma(t)$ with same parameter as in fig. 1, but with application of a DKC pulse after \SI{10}{ms} of free propagation. The inset plot shows the collimation process of the matter beam. \textbf{Right}: Contour lines depicting the difference in width $\sigma(t;b=0)-\sigma(t;b)$ for different $b$, of the same experiment using DKC as described in the left graph.}
\end{figure*}

As shown, decreasing the initial kinetic and GP interaction energy is advantageous when searching for deviations due to a logarithmic nonlinearity. 
One can think of many different ways of achieving this, for example preparing BECs with less density (smaller atom number $N$ or larger initial width $\sigma(0)$, however this might not be favorable as high densities are an important factor in Bose-Einstein condensation), smaller $a$ (BEC species with inherently smaller scatter length or via tuning of Feshbach resonances) and removing the BEC's kinetic energy after the initial expansion via external potentials which slow down the expansion.\\
\indent
The latter can be achieved, for example, using optical or magnetic delta kick collimation (DKC) techniques, in which a trapping potential is shortly switched on in order to counter the BEC's dispersion \cite{Ammann1997}. This magnetic/optical lens can ultimately be used to generate a collimated (minimally spreading) matter beam with low interaction energy.

\subsection{Enhancing the logarithmic nonlinearitiy's effects using delta kick collimation}

Using DKC we are able to diminish effects due to the high initial interaction energy and decrease $\chi$, which in turn, magnifies the logarithmic nonliearity's effects on the BEC's dynamics.
This has been similarly used, in order to infer bounds on collapse models from observation of collimated cold atom clouds \cite{Kovachy2015,BILARDELLO2016}.\\\indent
For the simulation of free expansion experiments with DKC, we choose the same parameters as in the previously discussed simulations of section \ref{sec:a}. 
The only difference being that after a time $t_\mathrm{DK}=\SI{10}{\milli\second}$, an harmonic DKC pulse is applied. 
The angular frequency $\omega$ of the harmonic potential (see equation \eqref{eq:gaussian-width-over-time}) is set to a value that leads to $\beta = \Dot{\sigma}= 0$ (infinite radius of curvature) at the end of the DKC sequence.
The DKC pulse is so short ($\Delta t = \SI{10}{\micro\second}$) that we approximate it as a thin lens.
This means that in order to collimate the matter beam $\omega^2 \Delta t \approx 1/t_\mathrm{DK}$ \cite{Kovachy2015}.
The simulated time evolution of a BEC's width $\sigma(t)$ subjected to this DKC pulse is shown in the left graph of figure 2.

We see that the collimation of the BEC does indeed increase the deviations if compared to the free expansion in fig. 1. The value of $\chi$ is in the order of unity for all shown examples.
The evolution under the current upper bound on $b$ (blue, dashed line) already differs by $\approx\SI{60}{\micro\meter}$ after \SI{1}{\second} of free propagation, instead of previously $\approx\SI{20}{\micro\meter}$ without DKC. The simulated expansions for larger values for $b$ (dotted, green, orange and red line) show the previously described behavior in the case of $\chi \ll 1$, where the BEC's maximum width is given by the prepared width using the DQK pulse.
\\
\indent
The difference in width $\sigma(t;b=0)-\sigma(t;b)$ for these experiments is shown in the left graph of figure 2. 
It becomes apparent that current experiments of BECs in microgravity could be able to lower the bound on $b$, since BEC's have already been generated with multiple seconds of free fall time \cite{vanZoest2010, Rudolph2011, Muntinga2013, Becker2018}. For example, the width differs by \SI{5}{\micro\meter} after \SI{1}{s} of free propagation, with a value of $b$, one magnitude lower than the current limit.
\\
\indent

\subsection{Error Estimation} \label{sec:error-estimation}
We give a brief error estimation of the proposed experiment.
\\
\indent
According to the GPE, the width's rate of change during free expansion of a BEC reaches a constant value (see eq. \eqref{eq:energy-per-particle}) in the far-field $t\gg m\sigma^2(0)/\hbar$ (also $t\gg t_\mathrm{DK}$, so we set $t_\mathrm{DK} = 0$ for convenient notation). 
The rate of expansion is
\begin{align}
    \Dot{\sigma}^2&(t\gg m\sigma^2(0)/\hbar)
    \nonumber\\
    &=\Dot{\sigma}^2(0)+ \frac{\hbar^2}{4 m^2}\sigma^{-2}(0) + \frac{\hbar^2 N a}{6\sqrt{\pi}m^2}\sigma^{-3}(0)  \label{eq:rate-of-expansion} \\
    &=\Dot{\sigma}_\mathrm{R}^2 + \Dot{\sigma}_\mathrm{HU}^2 + \Dot{\sigma}_\mathrm{GP}^2
    \nonumber
    .
\end{align}
Where the indices R, HU and GP indicate the individual contributions due to residual rate of change, Heisenberg uncertainty and Gross-Pitaevskii interaction respectively.
We can interpret our experiment as a test of this linear expansion in the far-field. 
Using formula \eqref{eq:rate-of-expansion}, we can estimate an error on the measurement of $\sigma(t)$ by acknowledging uncertainties of $N$, $a$, $\sigma(0)$ and $\Dot{\sigma}(0)$. 
The relative error $\delta_{\Dot{\sigma}(t)}=\Delta_{\Dot{\sigma}(t)}/\Dot{\sigma}(t)$ of the BEC's width can be estimated as
\begin{align}
    &\delta_{\Dot{\sigma}(t)} = \frac{1}{ \Dot{\sigma}_\mathrm{R}^2 + \Dot{\sigma}_\mathrm{HU}^2 + \Dot{\sigma}_\mathrm{GP}^2 }  \nonumber\\
    & \times \sqrt{
    \Dot{\sigma}_\mathrm{R}^4\delta_{\Dot{\sigma}(0)}^2
    +
    \Big(\frac{3}{2}\Dot{\sigma}_\mathrm{GP}^2 + \Dot{\sigma}_\mathrm{HU}^2 \Big)^2 \delta_{\sigma(0)}^2
    +
     \frac{1}{4}\Dot{\sigma}_\mathrm{GP}^4 \Big( \delta_a^2 + \delta_N^2 \Big)
    }
    ,
\end{align}
where $\delta_a$, $\delta_N$, $\delta_{\Dot{\sigma}(0)}$ and $\delta_{\sigma(0)}$ are the relative errors of $a$, $N$, $\Dot{\sigma}(0)$ and $\sigma(0)$ respectively.
The errors of $N$ and $a$ are on equal footings in terms of influence on $\delta_{\sigma(t)}$. However, we see that $\delta_{\sigma(0)}$ has an especially strong impact on the overall error. 
This is due to the cubic impact of $\sigma(0)$ on the interaction energy and the quadratic dependence of the quantum pressure.
\\
\indent
For example, the total estimated error $\Delta_{\Dot{\sigma}(t)}$ for the simulated experiment in fig. 2 is smaller than $\SI{0.1}{\micro\meter\per\second}$ if all relative errors are of 1\%. 
We see that \SI{60}{\micro\meter} is approximately the expected deviation in width after one second of free expansion with the current known bound. 
Even with a 20\% relative error in $N$, $a$ and $\sigma(0)$, the error is \SI{35}{\micro\meter\per\second} and thus, after one second of free propagation, smaller then the expected deviation.
\\
\indent
In addition to the exact characterization of the BEC's initial state which, in theory, could determine $\sigma(t)$, there are many environmental factors which could give rise to similar effects. 
For example, a inhomogeneous magnetic field $B$ might lead (in first order) to an effective harmonic potential of frequency $\omega_B$ that could lead to a narrowing. 
In the case of the Gaussian shaped wave function, the logarithmic nonlinearitity can basically seen as time-dependent harmonic potential. 
One can calculate that the effect of a logarithmic nonlinearity and a parasitic harmonic potential are distinguishable if $\omega_B < b/\hbar \approx \SI{1}{\per\second}$, with the current upper bound on $b$. 
This should be achievable if the BEC is prepared in a magnetic insensitive state, such that only quadratic Zeeman effects would need to be accounted for.\\
\indent
Furthermore, it would be important to distinguish a possible fundamental nonlinearity from an effective, nonlinear, logarithmic dynamic, for which there exist different ideas how those could arise \cite{Kostin1972, Schuch1983, Doebner1994, Nassar2013, Chavanis2018}.


\section{Summary and Discussion}

In this work the feasibility of tests for a logarithmic nonlinearity in the SE using Bose-Einstein condensates was examined.
Approximating the wave function in a Hartree ansatz, we proposed the logarithmic Gross-Pitaevskii equation which describes BECs governed by the LogSE.
We analyzed the free expansion of the logarithmic BEC with several hundreds of milliseconds of free fall time.
As these time scales become progressively more accessible 
\cite{vanZoest2010, Rudolph2011, Muntinga2013, Becker2018, Aguilera2014, Elliott2018, beccal2019},
new tests of the Schródinger equation and the resulting wave nature of matter can yield strengthened confirmation for the exactness of quantum mechanics.
However, under typical experimental parameters, the dynamics induced by two-body interactions largely outweigh those due to a possible nonlinearity.
Therefore, we proposed using optical or magnetic potentials to collimate the condensate to obtain a coherent-matter wave which would propagate as described by the SE.
In contrast to the linear free expansion, the LogSE spatially confines a wave-packet.
This effect can be measured by looking for a narrowing of the BEC's density distribution in the far-field.
We have shown that available free-expansion experiments can be used to determine new bounds on $b$ at least one order of magnitude below the current known value.
\\
\indent
It should be noted that the procedure we followed in order to derive the LogGPE, can be applied for every nonlinearity which respects the separability condition. 
There are several additional nonlinearities, usually proposed for dissipative and diffusion processes in open quantum systems, which have this property \cite{Kostin1972, Schuch1983, Doebner1994, Nassar2013, Chavanis2018} and could therefore be incorporated. 
This is especially helpful since, as mentioned in section \ref{sec:error-estimation}, it would be necessary to distinguish processes due to residual interactions with the environment from fundamental nonlinearities.
\\
\indent
The search for nonlinearities in the SE is of fundamental importance. 
Since Weinberg's proposal of a general nonlinear framework \cite{Weinberg1989} and its experimental testings \cite{Bollinger1989}, searches for deterministic deviations have, to our knowledge, completely ceased. 
This is indeed interesting, because since the writings of the first articles showing the possibility of superluminal signaling \cite{Gisin1990,Czachor1991,Polchinski1991} there have been several publications at least questioning its consequences \cite{Polchinski1991, Czachor1997, MIELNIK2001,Czachor2002, Kent2005, Jordan2010}.


\subsection*{Acknowledgments}
This work is supported by the German Space Agency DLR with funds provided by the Federal Ministry of Economics and Technology (BMWi) under grant number DLR50WP1432 and DLR50WM1852. \\\indent
We thank Simon Kanthak for fruitful discussions.

\bibliography{main}

\end{document}